\RequirePackage{fix-cm}
\documentclass[smallcondensed]{svjour3}     % onecolumn (ditto)
\smartqed  % flush right qed marks, e.g. at end of proof
\usepackage{graphicx}
\usepackage{amsmath}
\usepackage[square,sort,semicolon]{natbib}
\usepackage{color}
\newcommand{\mm}[1]{\textcolor{red}{#1}}
\begin{document}

\title{The linear Boltzmann equation in column experiments of porous media
}


\author{Kenji Amagai \and Motoko Yamakawa \and 
Manabu Machida \and Yuko Hatano
}


\institute{
K. Amagai \and M. Yamakawa \and Y. Hatano \at 
Graduate School of Systems and Information Engineering, 
University of Tsukuba, Tsukuba 305-7361, Japan\\
Y. Hatano \email{hatano@risk.tsukuba.ac.jp}\\
M. Machida \at
Institute for Medical Photonics Research, Hamamatsu University School of Medicine, Hamamatsu 431-3192, Japan
}

\date{Received: date / Accepted: date}

\maketitle

\begin{abstract}
The use of the linear Boltzmann equation is proposed for transport in porous media in a column. By column experiments, we show that the breakthrough curve is reproduced by the linear Boltzmann equation. The advection-diffusion equation is derived from the linear Boltzmann equation in the asymptotic limit of large propagation distance and long time.
\end{abstract}

\section{Introduction}
\label{intro}

Mass transport in porous media is one of the central topics in hydrology. For example, mass transport has been intensively studied in the contexts of carbon dioxide capture and storage (\citealt{Benson_2008}) and enhanced oil recovery (\citealt{Thomas_2008}). Moreover migration of the soil pollution of radioactive cesium in Fukushima, Japan, which is considered as mass transport in porous media, is an urgent issue (\citealt{Matsuda_2015}). The phenomenon of mass transport in porous media has been analyzed by the advection-diffusion equation (ADE) since 1950s. However, different examples in which ADE is not applicable have been found. Adams and Gelhar showed that the mass transport predicted by ADE was significantly different from measured values in a field scale experiment (\citealt{Adams_1992}). In the laboratory scale, Berkowitz, Scher, and Silliman found that the transport in a laboratory flow cell does not obey ADE, but it shows anomalous diffusion (or non-Fickian transport) and deviates from Fick's law (\citealt{Berkowitz_2000}). In such flow of anomalous diffusion, unlike the prediction by ADE, the peak of concentration barely moves and its distribution has a long tail. The deviation from ADE was investigated by a random particle simulation (\citealt{Kennedy_2001}).

To reproduce anomalous diffusion, continuous time random walk (CTRW) has been developed since the first proposal by Montroll and Weiss (\citealt{Montroll_1965}). Anomalous diffusion in porous media in different situations can be reproduced by CTRW when waiting time and jump distance are suitably adjusted. Although CTRW is widely used in the study of porous media (\citealt{Berkowitz_1998,Hatano_1998,Levy_2003,Nissan_2017}), the relation between the movement of tracer particles and the waiting time and jump distance of CTRW is not apparent. By taking the correlation in waiting time into account, correlated continuous time random walk was proposed (\citealt{Montero_2007}). In addition to CTRW, the following models were proposed. In tempered anomalous diffusion, strong anomalous diffusivity is suppressed by the introduction of an exponential cutoff for waiting times (\citealt{Meerschaert_2008}). The stochastic hydrology was developed (\citealt{Gelhar_1986,Rubin_2003}). The mobile/immobile (MIM) model and fractional MIM model divide mass transport into the moving part and immovable part (\citealt{Genuchten_1976,Schumer_2003}). It is known that diffusion equations with fractional derivatives are obtained in the asymptotic limit of CTRW (\citealt{Metzler_2000}). In fractional advection-dispersion equation,  time and spatial derivatives are replaced by fractional derivatives (\citealt{Benson_2000}). The use of fractional derivatives has attracted attention as a theoretical tool for anomalous diffusion. Fractional derivative models are compared (\citealt{Wei_2016,Sun_2017}), parameters in fractional equations were estimated (\citealt{Chakraborty_2009,Kelly_2017}), the physical meaning of fractional derivatives was discussed (\citealt{Liang_2019}), and orders of fractional derivatives were taken to be variables (\citealt{Sun_2009}).

To study transport in porous media, in addition to the concentration in mass transport, actual flow was visualized (\citealt{Moroni_2009,Sen_2012}) and the flow was reproduced by large scale simulation (\citealt{Anna_2013,Liu_2016}). These early studies show that tracer particles and fluid particles change velocity as a function of time when along the flow path they make a detour around obstacles in the the porous medium.

\begin{figure}
\centering
\includegraphics[clip,width=0.4\textwidth]{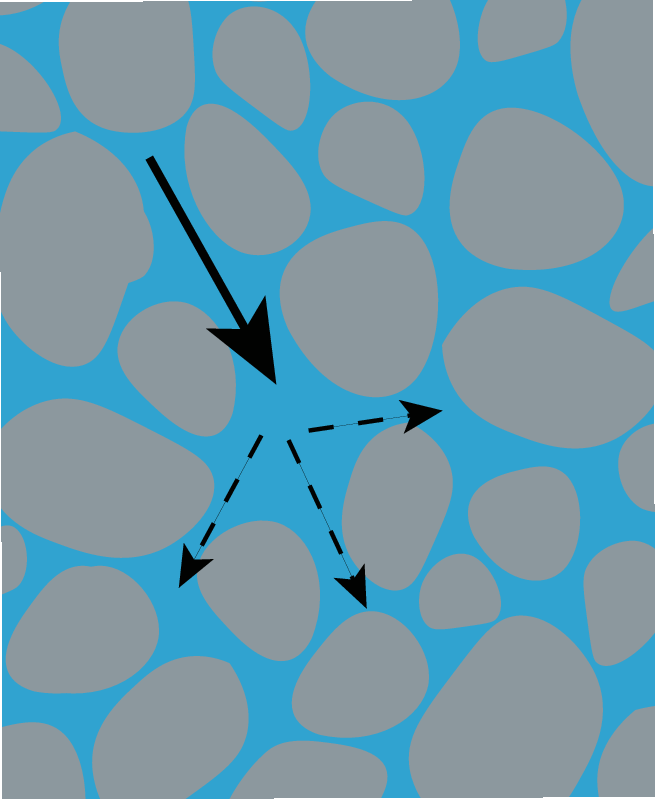}
\caption{
(Color online) The transport of tracer particles in a porous medium. This transport is regarded as noninteracting particles which undergo scatterings by scatterers.
}
\label{fig:schematic}
\end{figure}

Although CTRW and fractional equations have achieved certain success, the physical meaning of the random walk process and physical background of fractional-order derivatives are not easy to understand. In this paper, we propose to use the linear Boltzmann equation (LBE), in which the physical scattering process is more apparent. Figure \ref{fig:schematic} illustrates the Boltzmann transport picture of the transport in porous media. Tracer particles in a porous medium flow along fissures and fractures. We regard the process of a tracer particle changing its direction as a scattering. Thus for the ensemble of tracer particles, the transport in the porous medium can be viewed as the transport of noninteracting particles which are scattered by scatterers in different directions with a certain probability distribution.

The rest of the paper is organized as follows. In Sec.~\ref{sec:Method}, the linear Boltzmann equation is introduced. We explore the advection-diffusion approximation in Sec.~\ref{sec:ADA}. In Sec.~\ref{sec:ADO}, we solve the linear Boltzmann equation. Section \ref{sec:column} is devoted to column experiments. Conclusions are given in Sec.~\ref{sec:concl}.

\section{Linear Boltzmann transport}
\label{sec:Method}

The linear Boltzmann equation, radiative transport equation, or transport equation governs the transport of noninteracting particles with scattering and absorption. The equation describes neutron transport in a reactor (\citealt{Case_1967}) and light transport in random media such as clouds and the intersteller medium (\citealt{Apresyan_1996,Chandrasekhar_1960,Ishimaru_1978,Sobolev_1976,Thomas_1999}). Let us regard the transport of tracer particles along flow paths as the transport of particles in a homogeneous random medium. See the schematic figure in Fig.~\ref{fig:schematic}. There is a random arrangement of flow paths in beads and sand in column experiments. At a branching point where several paths split into different directions, tracer particles are {\em scattered} to these directions with certain probabilities. The average distance that a tracer particle travels along a flow path between branching points is an effective mean free path $v_0/\sigma_s$ (see (\ref{proposed_model}) for $v_0,\sigma_s$). Indeed, the use of the linear Boltzmann equation for the flow of tracer particles in porous media was proposed by Williams (\citealt{Williams_1992a,Williams_1992b,Williams_1993a,Williams_1993b}).

A remark is necessary for our Boltzmann transport. The (nonlinear) Boltzmann equation can be simulated by the lattice Boltzmann scheme (\citealt{Succi01}), and the lattice Boltzmann method has become one of commonly used numerical methods to investigate the flow in porous media. Instead of fluid in porous media, we focus on the Boltzmann transport of tracer particles in the fluid. Hence, the proposed equation (\ref{proposed_model}) below is a linear equation.

Having in mind the transport in column experiments which is described in Sec.~\ref{sec:column}, we consider the one-dimensional linear Boltzmann equation. Let $\mu\in[-1,1]$ be the cosine of the polar angle and $v_0>0$ be the inherent particle speed. Although the coefficient of the spatial derivative is $v_0\mu$ in Williams' study (\citealt{Williams_1992a,Williams_1992b}) , in this paper, we take into account advection, which will be denoted by $u\ge0$. Hence the velocity $v$ is written as $v=u+v_0\mu$. Let $\psi(x,\mu,t)$ be the angularly dependent number density of particles at position $x\ge0$ in direction $\mu$ at time $t\ge0$. We can write the linear Boltzmann equation as follows.
\begin{equation}
\label{proposed_model}
\frac{\partial}{\partial t}\psi(x,\mu,t)+(u+v_0\mu)\frac{\partial}{\partial x}\psi+(\sigma_a+\sigma_s)\psi=
\frac{\sigma_s}{2}\int^{1}_{-1}\psi(x,\mu,t)\,\mathrm{d}\mu.
\end{equation}
where isotropic scattering is assumed in the integral on the right-hand side. Here, $\sigma_a\ge0$ and $\sigma_s>0$ are the absorption and scattering coefficients, respectively. Let $n_0$ be the initial particle number density. Tracer particles are injected to the column at $x=0$ in the normal direction. The initial and boundary conditions are given by
\begin{eqnarray}
\label{analytical solution1_1}
&&\psi(x,\mu,0)=0,\\
\label{analytical solution1_2}
&&\psi(0,\mu,t)=n_0\delta(\mu-1),\quad\mu>-\eta,
\end{eqnarray}
where $\delta(\mu-1)$ is the Dirac delta function and
\begin{equation}
\label{analytical solution1_8}
\eta=\frac{u}{v_0}.
\end{equation}
We note that unlike the usual half-range boundary condition of $\mu>0$ (\citealt{Case_1967}), the boundary value in (\ref{analytical solution1_2}) is specified for $\mu\in(-\eta,1]$ due to the presence of the advection. We regard the column as the half space ($0<x<\infty$) and have $\psi\to0$ as $x\to\infty$.

The particle number density $n(x,t)$ is calculated as
\begin{equation}
\label{defnumdens}
n(x,t)=\int_{-1}^1\psi(x,\mu,t)\,\mathrm{d}\mu.
\end{equation}
Then the concentration is given by
\begin{equation}
C(x,t)=\alpha n(x,t),
\end{equation}
where $\alpha$ is the mass per particle.

\section{Advection diffusion approximation}
\label{sec:ADA}

Let us derive the advection-diffusion equation from the linear Boltzmann equation. According the usual procedure of the diffusion approximation (\citealt{Duderstadt_1979,Ishimaru_1978}), we assume that the angular dependence of $\psi(x,\mu,t)$ is weak and written as 
\begin{equation}
\label{LBE_ADE_1}
\psi(x,\mu,t)=\frac{1}{2\alpha}C(x,t)+\frac{3}{2\alpha v_0}J(x,t)\mu,
\end{equation}
where,
\begin{align}
\label{LBE_ADE_2}
C(x,t)&=
\alpha\int^{1}_{-1}\psi(x,\mu,t)\,\mathrm{d}\mu,
\\
\label{LBE_ADE_3}
J(x,t)&=
\alpha v_0\int^{1}_{-1}\mu\psi(x,\mu,t)\,\mathrm{d}\mu.
\end{align}
In (\ref{LBE_ADE_1}), we implicitly assumed a large propagation distance and long time. If absorption is strong, $\psi$ becomes zero before the form (\ref{LBE_ADE_1}) is achieved. Hence we assume $\sigma_a$ is small. Indeed in Sec.~\ref{sec:column}, we will see that $\sigma_a$ is negligibly small. Indeed, the approximation in (\ref{LBE_ADE_1}) is called the $P_1$ approximation (\citealt{Case_1967}). Below, we will proceed further and obtain the diffusion equation. We substitute the form (\ref{LBE_ADE_1}) for $\psi$ in (\ref{proposed_model}). By integrating the resulting equation over $\mu$, we obtain
\begin{equation}
\label{LBE_ADE_4}
\frac{\partial C}{\partial t}+u\frac{\partial C}{\partial x}+\frac{\partial J}{\partial x}+\sigma_aC=0.
\end{equation}
By integrating the equation over $\mu$ after multiplying $\mu$, we obtain
\begin{equation}
\label{LBE_ADE_5}
\frac{1}{v_0}\frac{\partial J}{\partial t}+\frac{u}{v_0}\frac{\partial J}{\partial x}+\frac{v_0}{3}\frac{\partial C}{\partial x}+
\frac{\sigma_a+\sigma_s}{v_0}J=0.
\end{equation}
We assume that the first term and second term of the above equation are small and can be dropped. Then we have
\begin{equation}
\label{LBE_ADE_6}
J=-D\frac{\partial C}{\partial x},
\end{equation}
where
\begin{align}
\label{LBE_ADE_8}
D&=\frac{v_0\ell^*}{3},
\\
\label{LBE_ADE_9}
\ell^*&=\frac{v_0}{\sigma_a+\sigma_s}\simeq\frac{v_0}{\sigma_s}.
\end{align}
Hence we obtain the following advection diffusion equation by substituting (\ref{LBE_ADE_6}) into (\ref{LBE_ADE_4}).
\begin{equation}
\label{LBE_ADE_7}
\frac{\partial C}{\partial t}-D\frac{\partial^2 C}{\partial x^2}+u\frac{\partial C}{\partial x}+\sigma_aC=0,
\end{equation}
For the photon diffusion equation, $\ell^*$ is known to be independent of the absorption coefficient (\citealt{Furutsu_1994}). The diffusion approximation holds when $x$ is sufficiently larger than the transport mean free path $\ell^*$.

The diffusion equation (\ref{LBE_ADE_7}) reduces to the usual advection-diffusion equation when $\sigma_a=0$. Since we found $\sigma_a\approx0$ in our column experiments, the solution by Ogata and Banks (\citealt{Ogata_1961}) is used in this paper when the concentration calculated from the linear Boltzmann equation is compared to the advection-diffusion equation. In Appendix \ref{wabsorption}, the solution to (\ref{LBE_ADE_7}) is presented.

Finally, we consider the boundary condition. Let us substitute (\ref{LBE_ADE_1}) for $\psi(x,\mu,t)$ in the boundary condition (\ref{analytical solution1_2}). By integrating both sides of the resulting equation over $\mu$ from $-\eta$ to $1$, we obtain
\begin{equation}
\label{bc2}
C(0,t)-\frac{3D\eta(1-\eta)}{2}\frac{D}{u}\frac{\partial C}{\partial x}(0,t)=\frac{2\alpha n_0}{1+\eta}.
\end{equation}
Thus we obtain the Robin boundary condition. In this paper, we neglect the flux term and write
\begin{equation}
\label{bc1}
C(0,t)=C_0,
\end{equation}
where $C_0>0$ denotes the boundary source. Since the diffusion approximation does not hold on the boundary, the actual proportionality constant between $\alpha n_0$ and $C_0$ is a fitting parameter. We introduce
\begin{equation}
\beta=\frac{\alpha n_0}{C_0}.
\end{equation}

\section{Concentration from the linear Boltzmann equation}
\label{sec:ADO}

Let us consider the Laplace transform.
\begin{equation}
\label{analytical_solution_1_5}
\hat{\psi}(x,\mu,p)=\int^{\infty}_0\mathrm{e}^{-pt}\psi(x,\mu,t)\,\mathrm{d}t.
\end{equation}
We introduce
\begin{equation}
\label{analytical solution1_6}
\mu_t=\frac{\sigma_a+\sigma_s+p}{v_0},\quad
\mu_s=\frac{\sigma_s}{v_0}.
\end{equation}
Then our transport equation is written as
\begin{equation}
\label{analytical_solution_1_4}
(\eta+\mu)\frac{\partial}{\partial x}\hat{\psi}(x,\mu,p)+\mu_t\hat{\psi}(x,\mu,p)=\frac{\mu_s}{2}\int^{1}_{-1}\hat{\psi}(x,\mu,p)\,\mathrm{d}\mu,
\end{equation}
with the boundary condition,
\begin{equation}
\label{analytical solution1_9}
\hat{\psi}(0,\mu,p)=\frac{n_0}{p}\delta(\mu-1),\quad\mu>-\eta.
\end{equation}

We will compute $\hat{\psi}$ with the analytical discrete ordinates method \mm{(ADO)} (\citealt{Siewert_1999,Barichello_2000,Barichello_2001}). 
Unlike the standard ADO, the coefficient $\mu_t$ is a complex number since the variable $p$ is complex and moreover, the boundary condition is given for $\mu>-\eta$ instead of $\mu>0$. 
We begin by decomposing $\hat{\psi}$ into the ballistic term $\hat{\psi}_b$ and scattering term $\hat{\psi}_s$ as
\begin{equation}
\label{analytical solution2_3}
\hat{\psi}(x,\mu,p)=\hat{\psi_b}(x,\mu,p)+\hat{\psi_s}(x,\mu,p).
\end{equation}
Here, $\hat{\psi}_b$ obeys
\begin{equation}
\label{analytical solution2_4}
(\eta+\mu)\frac{\partial}{\partial x}\hat{\psi_b}+\mu_t\hat{\psi_b}=0,
\quad x>0,\;-1\le\mu\le 1,
\end{equation}
with the boundary condition
\begin{equation}
\hat{\psi_b}(0,\mu,p)=\frac{n_0}{p}\delta(\mu-1),
\quad\mu>-\eta,
\end{equation}
and $\hat{\psi}_s$ obeys 
\begin{equation}
\label{analytical solution2_7}
(\eta+\mu)\frac{\partial}{\partial x}\hat{\psi_s}+\mu_t\hat{\psi_s}=\frac{\mu_s}{2}\int^{1}_{-1}\hat{\psi_s}\,\mathrm{d}\mu+q,
\quad x>0,\;-1\le\mu\le1,
\end{equation}
with the boundary condition
\begin{equation}
\hat{\psi_s}(0,\mu,p)=0,
\quad\mu>-\eta,
\end{equation}
where
\begin{equation}
\label{analytical solution2_9}
q(x,\mu,p)=\frac{\mu_s}{2}\int^{1}_{-1}\hat{\psi_b}(x,\mu,p)\,\mathrm{d}\mu=
\frac{n_0\mu_s}{2p}\mathrm{e}^{-x\mu_t/(\eta+\mu)}.
\end{equation}
We obtain
\begin{equation}
\label{analytical solution2_6}
\hat{\psi_b}(x,\mu,p)=
\frac{n_0}{p}\mathrm{e}^{-x\mu_t/(\eta+\mu)}\delta(\mu-1).
\end{equation}

Let us drop $p$ and write $\hat{\psi}_s(x,\mu)=\hat{\psi}_s(x,\mu,p)$ and $q(x,\mu)=q(x,\mu,p)$. For the computation of $\hat{\psi}_s$, we discretize the integral by the Gauss-Legendre quadrature and obtain
\begin{align}
\label{analytical solution3_5}
&
(\eta+\mu_i)\frac{\partial}{\partial x}\hat{\psi_s}(x,\mu_i)+
\mu_t\hat{\psi_s}(x,\mu_i)
\nonumber \\
&=
\frac{\mu_s}{2}\sum^{N}_{j=1}w_j
\left[\hat{\psi_s}(x,\mu_j)+\hat{\psi_s}(x,-\mu_j)\right]+q(x,\mu_i),
\end{align}
where $\mu_i,w_i$ ($i=1,2,\dots,2N$) are abscissas and weights, respectively. These $\mu_i,w_i$ with $0<\mu_1<\cdots<\mu_N<1$ and $\mu_{N+i}=-\mu_i$ ($i=1,\dots,N$) can be calculated by the Golub-Welsch algorithm (\citealt{Golub_1969}). In the numerical calculation in Sec.~\ref{sec:column}, we set
\begin{equation}
N=30.
\end{equation}
Furthermore, we introduce $N_{\eta}$ as the largest integer such that $-\eta<\mu_{N_{\eta}}$. The scattering part $\hat{\psi}_s$ is obtained as
\begin{equation}
\label{analytical solution3_7}
\hat{\psi_s}(x,\mu_i)=
\sum^{2N}_{j=1}\int^{\infty}_{0}G(x,\mu_i;x',\mu_j)q(x',\mu_j)\,\mathrm{d}x',
\end{equation}
where the Green's function defined for each $p$ satisfies
\begin{align}
\label{analytical solution3_8}
(\eta+\mu_i)\frac{\partial}{\partial x}G(x,\mu_i;x_0,\mu_{i_0})+
\mu_tG(x,\mu_i;x_0,\mu_{i_0})
&=
\frac{\mu_s}{2}\sum^{2N}_{j=1}w_jG(x,\mu_j;x_0,\mu_{i_0})
\nonumber \\
&+
\delta(x-x_0)\delta_{ii_0},
\end{align}
where $\delta_{ii_0}$ is the Kronecker delta. Here the boundary condition is given by
\begin{equation}
G(0,\mu_i;x_0,\mu_{i_0})=0,\quad\mu_i>\mu_{N_{\eta}}.
\end{equation}

Let us consider the following homogeneous equation to calculate the Green's function with ADO (\citealt{Siewert_1999,Barichello_2000,Barichello_2001}). Note that the equation will be solved for each $p$.
\begin{equation}
\label{analytical solution4_1}
\left((\eta+\mu_i)\frac{\partial}{\partial x}+\mu_t\right)\hat{\psi}(x,\mu_i)=\frac{\mu_s}{2}\sum^{2N}_{j=1}w_j\hat{\psi}(x,\mu_j).
\end{equation}
We note that $\hat{\psi}$ depends on $p$ through $\mu_t$. With separation of variables, we can write $\hat{\psi}$ as
\begin{equation}
\label{analytical solution4_2}
\hat{\psi}(x,\mu_i)=\phi(\nu,\mu_i)\mathrm{e}^{-x/\nu},
\end{equation}
where $\nu$ is the separation constant. The function $\phi(\nu,\mu_i)$ satisfies the normalization condition,
\begin{equation}
\label{analytical solution4_3}
\sum^{2N}_{i=1}w_i\phi(\nu,\mu_i)=
\sum^{N}_{i=1}w_i\left(\phi(\nu,\mu_i)+\phi(\nu,-\mu_i)\right)=1.
\end{equation}
For the integral equation in which the sum on the right-hand side of (\ref{analytical solution4_1}) is replaced by the integral $\frac{\mu_s}{2}\int_{-1}^1\hat{\psi}\,\mathrm{d}\mu$, this $\phi$ is called the singular eigenfunction (\citealt{Case_1960}). However, by the discretization, we obtain
\begin{equation}
\label{analytic_solution_4_5}
\phi(\nu,\mu_i)=\frac{\mu_s\nu}{2}\frac{1}{\mu_t\nu-\mu_i-\eta}.
\end{equation}
assuming $\nu\neq(\mu_i+\eta)/\mu_t$. It is known that $\mu\neq\mu_i/\mu_t$ if $\eta=0$ and $p$ is real (\citealt{Siewert_1999}). We can show the following orthogonality relation.
\begin{equation}
\label{analytic_solution_4_8}
\sum^{2N}_{i=1}w_i(\mu_i+\eta)\phi(\nu,\mu_i)\phi(\nu',\mu_i)=
\mathcal{N}(\nu)\delta_{\nu\nu'},
\end{equation}
where
\begin{equation}
\label{analytic_solution_4_9}
\mathcal{N}(\nu)=
\sum^{2N}_{i=1}w_i(\mu_i+\eta)\phi(\nu,\mu_i)^2.
\end{equation}
We can find $2N$ eigenvalues $\nu=\nu_n$ ($n=1,2,\dots,2N$). Moreover there are $N_{\eta}$ eigenvalues with positive real parts. See Appendix \ref{eigen} for the eigenvalues.

Taking the fact that $G$ vanishes as $x\to\infty$, we can write
\begin{align}
\label{analytic_solution_5_1}
G(x,\mu_i;x_0,\mu_{i_0})
&=
G_0(x,\mu_i;x_0,\mu_{i_0})
\nonumber \\
&+
\sum_{\Re{\nu_n}>0}B(\nu;x_0,\mu_{i_0})\phi(\nu,\mu_i)\mathrm{e}^{-x/\nu},
\end{align}
where coefficients $B(\nu;x_0,\mu_{i_0})$ will be determined later. The free-space Green's function $G_0$ is calculated as
\begin{equation}
\label{analytic_solution_5_6}
G_0(x,\mu_i;x_0,\mu_{i_0})=
\pm\sum_{\pm\Re{\nu_n}>0}\frac{w_{i_0}}{\mathcal{N}(\nu_n)}
\phi(\nu_n,\mu_{i_0})\phi(\nu_n,\mu_i)\mathrm{e}^{-(x-x_0)/\nu_n},
\end{equation}
where upper signs are chosen for $x>x_0$ and lower signs are used for $x<x_0$. Since $G$ vanishes at $x=0$ for $\mu_i>\mu_{N_{\eta}}$, we have from (\ref{analytic_solution_5_1}),
\begin{equation}
\label{analytic_solution_5_7}
\sum_{\Re{\nu_n}>0}B(\nu;x_0,\mu_{i_0})\phi(\nu_n,\mu_i)=
\sum_{\Re{\nu_n}<0}\frac{w_{i_0}}{\mathcal{N}(\nu_n)}
\phi(\nu_n,\mu_{i_0})\phi(\nu_n,\mu_i)\mathrm{e}^{x_0/\nu_n}.
\end{equation}
Let us multiply (\ref{analytic_solution_5_7}) by $\exp(-x_0\mu_t/(\eta+1))$, then integrate both sides with $x_0$ and take the sum with respect to $i$. We obtain
\begin{equation}
\label{analytic_solution_6_1}
\sum_{\Re{\nu_n}>0}E(\nu_n)\phi(\nu_n,\mu_i)=
\sum_{\Re{\nu_n}<0}
\frac{-\nu_n(\eta+1)}{\mathcal{N}(\nu_n)(\eta+1-\nu_n\mu_t)}\phi(\nu_n,\mu_i),
\end{equation}
where
\begin{equation}
\label{analytic_solution_5_9}
E(\nu_n)=\sum_{i_0=1}^{2N}\int_0^{\infty}B(\nu_n;x_0,\mu_{i_0})
\mathrm{e}^{-\mu_tx_0/(\eta+1)}\,\mathrm{d}x_0.
\end{equation}
We can numerically obtain $E(\nu_n)$ from the linear system (\ref{analytic_solution_6_1}). Finally, we obtain
\begin{align}
\label{analytic_solution_6_3}
\hat{\psi_s}(x,\mu_i)
&=
\frac{n_0\mu_s}{2p}\sum_{\Re{\nu_n}>0}\left[
\frac{\nu_n}{\mathcal{N}(\nu_n)}\phi(\nu_n,\mu_i)
\frac{\eta+1}{\eta+1-\nu_n\mu_t}
\left(\mathrm{e}^{-\mu_tx/(\eta+1)}-\mathrm{e}^{-x/\nu_n}\right)
\right]
\nonumber \\
&+
\frac{n_0\mu_s}{2p}\sum_{\Re{\nu_n}<0}\left[
\frac{\nu_n}{\mathcal{N}(\nu_n)}\phi(\nu_n,\mu_i)
\frac{\eta+1}{\eta+1-\nu_n\mu_t}\mathrm{e}^{-\mu_tx/(\eta+1)}
\right]
\nonumber\\
&+
\frac{n_0\mu_s}{2p}\sum_{\Re{\nu_n}>0}E(\nu_n)\phi(\nu_n,\mu_i)
\mathrm{e}^{-x/\nu_n}.
\end{align}
In this way, $\hat{\psi}_s$ is computed using ADO.

The Laplace transform of the particle number density $\hat{n}(x,p)$ is obtained as
\begin{align}
\label{analytic_solution_6_4}
\hat{n}(x,p)&=
\int^1_{-1}\hat{\psi}(x,\mu,p)\,\mathrm{d}\mu
\nonumber\\
&=
\int^1_{-1}\hat{\psi_b}(x,\mu,p)\,\mathrm{d}\mu+
\int^1_{-1}\hat{\psi_s}(x,\mu,p)\,\mathrm{d}\mu
\nonumber\\
&=\frac{n_0}{p}\mathrm{e}^{-x\mu_t/(\eta+1)}
\nonumber\\
&+
\frac{n_0\mu_s}{2p}\sum_{\Re{\nu_n}>0}\left[
\frac{\nu_n}{\mathcal{N}(\nu_n)}\frac{\eta+1}{\eta+1-\nu_n\mu_t}\left(
\mathrm{e}^{-x\mu_t/(\eta+1)}-\mathrm{e}^{-x/\nu_n}\right)\right]
\nonumber\\
&+
\frac{n_0\mu_s}{2p}\sum_{\Re{\nu_n}<0}\left[
\frac{\nu_n}{\mathcal{N}(\nu_n)}\frac{\eta+1}{\eta+1-\nu_n\mu_t}
\mathrm{e}^{-x\mu_t/(\eta+1)}\right]
\nonumber\\
&+
\frac{n_0\mu_s}{2p}\sum_{\Re{\nu_n}>0}E(\nu_n)\mathrm{e}^{-x/\nu_n}.
\end{align}
We note that the particle number density $n(x,t)$ is given by
\begin{equation}
n(x,t)=
\frac{1}{2\pi\mathrm{i}}\int^{\gamma+\mathrm{i}\infty}_{\gamma-\mathrm{i}\infty}\mathrm{e}^{pt}\hat{n}(x,p)\,\mathrm{d}p.
\end{equation}
The Bromwich integral in the inverse Laplace transform is numerically evaluated by the trapezoidal rule. In general, deformation of the contour can be considered for the inversion (\citealt{Weideman_2007}). We found $\gamma=0.04$ is suitable. Hence, the concentration is obtained as
\begin{align}
C(x,t)
&=
\frac{\alpha n_0\mu_s}{4\pi\mathrm{i}}\int^{\gamma+\mathrm{i}\infty}_{\gamma-\mathrm{i}\infty}\frac{\mathrm{e}^{pt}}{p}
\Biggl\{\frac{2}{\mu_s}\mathrm{e}^{-x\mu_t/(\eta+1)}
\nonumber\\
&+
\sum_{\Re{\nu_n}>0}\left[
\frac{\nu_n}{\mathcal{N}(\nu_n)}\frac{\eta+1}{\eta+1-\nu_n\mu_t}\left(
\mathrm{e}^{-x\mu_t/(\eta+1)}-\mathrm{e}^{-x/\nu_n}\right)\right]
\nonumber\\
&+
\sum_{\Re{\nu_n}<0}\left[
\frac{\nu_n}{\mathcal{N}(\nu_n)}\frac{\eta+1}{\eta+1-\nu_n\mu_t}
\mathrm{e}^{-x\mu_t/(\eta+1)}\right]
\nonumber\\
&+
\sum_{\Re{\nu_n}>0}E(\nu_n)\mathrm{e}^{-x/\nu_n}
\Biggr\}\,\mathrm{d}p.
\end{align}

%\clearpage
\section{Column experiments}
\label{sec:column}

Column experiments are often used for the study of transport in random media (\citealt{Cotris_2004}). We here present results of a series of tracer breakthrough experiments conducted in a one-dimensional flow field. Fluid with tracer particles is injected by the peristaltic pump (MP-1000, Eyela). The flow rate of injected tracer solution is controlled by the peristaltic pump. Ultra pure water is used. First, water is injected for 24 hours or more until steady flow field is achieved. Next, tracer solution is injected to displace the fresh water. The discharged solution is collected by the fraction collector (CHF161RA, Advantec) at regular intervals. Column experiments are performed at room temperature of $25$--$26\,{}^{\circ}{\rm C}$.

Tracer experiments are conducted until the discharge concentration becomes equal to the influent concentration. For the sake of the accuracy of measurements, we correct the measurement time error from the tube volume and inlet volume of the column by subtracting the time lag due to the switchover from the breakthrough time.

\subsection{Column experiments with non-adsorbed solute}

First we use non-adsorbed solute. We use two different sizes of glass beads made with Soda glass; (Run 1) smaller beads with diameter ranging from $0.177\,{\rm mm}$ to $0.250\,{\rm mm}$ and (Run 2) larger beads with diameter from $0.500\,{\rm mm}$ to $0.710\,{\rm mm}$. The column length is $20.0\,{\rm cm}$ with the internal diameter of the column $2.9\,{\rm cm}$. In order to prevent entrainment of air, we carry out beads packing under saturated conditions, in which the water surface is kept at a constant height from the top of the beads layer. We stirrer beads while filling in order to remove small bubbles attached to the beads.

The inlet and the outlet ends of the column are separated from the porous medium of a glass filter with the pore size $10\,{\rm \mu m}$, originally for the purpose of the liquid chromatography. The glass filter serves to enhance tracer mixing and to dampen small flow pulses on injection. The tracer concentration is measured by the EC meter (LAQUAtwin-EC-33B, Horiba). The precision of the meter is $\pm 2\%$. The temperature compensation circuit is installed.

We prepare the NaCl solution of $0.1\%$ with the salt of purity $99.5\%$ (Wako). Injection of the solution is done by the peristaltic pump. The measured EC values are given by the unit of Sv/cm, we made the calibration curve and obtain the concentration.

\subsubsection{Run 1: small beads}
\label{exp_run1}

The column is filled with small beads. The measured porosity is $\eta=0.386$. The bed height is $18.0\,{\rm cm}$ with section area $6.61\,{\rm cm}^2$. The flow rate is $0.958\,{\rm cm}^3/{\rm min}$. 

\begin{figure}
\centering
\includegraphics[width=0.8\textwidth]{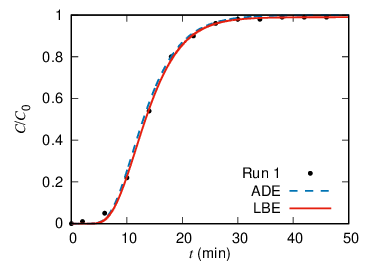}
\caption{
(Color online) The relative concentration $C/C_0$ for Run 1 (black dots) is plotted with calculated values from LBE (red solid line) and ADE (blue dashed line).
}
\label{fig:run1}
\end{figure}

\begin{figure}
\centering
\includegraphics[clip,width=0.8\textwidth]{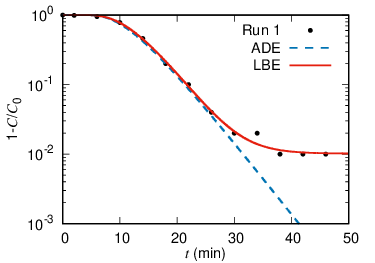}

\caption{
(Color online) Same as Fig.~\ref{fig:run1} but $1-C/C_0$ is plotted in the semi-log graph.
}
\label{fig:run1_log}
\end{figure}

Figure \ref{fig:run1} shows the chloride breakthrough curve from Run 1. The vertical axis shows the relative concentration $C/C_0$ and the horizontal axis is time $t$. The measured values from the column experiment (black dots) are compared to $C/C_0$ calculated by the linear Boltzmann equation (LBE) (red solid line) and by the ADE model (\citealt{Ogata_1961}) (blue dashed line). The parameters in both equations are determined by nonlinear least-squares fitting (i.e., the minimization of the sum of squares of the residuals between the experimental and theoretical curves); they are obtained as $\sigma_a=1.0\times 10^{-8}\,{\rm min}^{-1}$, $\sigma_s=5.1645\,{\rm min}^{-1}$, $v_0=5.3073\,{\rm cm}/{\rm min}$, $u=1.6445\,{\rm cm}/{\rm min}$, $\beta=0.09130$ for LBE, and $u=1.2886\,{\rm cm}/{\rm min}$, $D=1.8379\,{\rm cm}^2/{\rm min}$ for ADE. In Fig.~\ref{fig:run1}, experimental results agree well with both LBE and ADE. Figure \ref{fig:run1_log} shows semi-log plots of $1-C/C_0$ as a function of $t$.

\subsubsection{Run 2: large beads}
\label{exp_run2}

The column is filled with large beads. The measured porosity is $\eta=0.367$. (This value of the porosity is smaller than that of Run 1. This may be due to small bubbles attached on the beads surface during Run 1.) The bed height is $18.5\,{\rm cm}$ with section area $6.61\,{\rm cm}^2$. The flow rate is $0.875\,{\rm cm}^3/{\rm min}$.

\begin{figure}
\centering
\includegraphics[clip,width=0.8\textwidth]{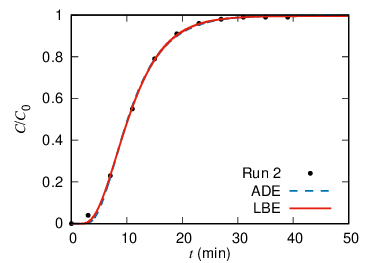}
\caption{
(Color online) The relative concentration $C/C_0$ for Run 2 (black dots) is plotted with calculated values from LBE (red solid line), and ADE (blue dashed line).
}
\label{fig:run2}
\end{figure}

\begin{figure}
\centering
\includegraphics[clip,width=0.8\textwidth]{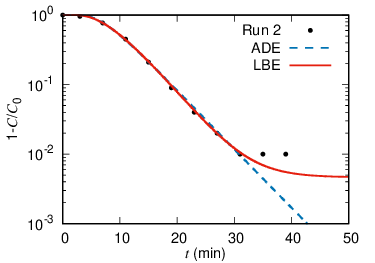}
\caption{
(Color online) Same as Fig.~\ref{fig:run2} but $1-C/C_0$ is plotted in the semi-log graph.
}
\label{fig:run2_log}
\end{figure}

Figure \ref{fig:run2} for Run 2 is the same as Fig.~\ref{fig:run1} but large glass beads are used. By nonlinear least-squares fitting, the parameters are obtained as $\sigma_a=1.0\times 10^{-8}\,{\rm min}^{-1}$, $\sigma_s=4.6778\,{\rm min}^{-1}$, $v_0=8.1753\,{\rm cm}/{\rm min}$, $u=2.4999\,{\rm cm}/{\rm min}$, $\beta=0.08579$ for LBE, and $u=1.6189\,{\rm cm}/{\rm min}$, $D=3.8519\,{\rm cm}^2/{\rm min}$ for ADE. In Fig.~\ref{fig:run2}, experimental results are described well by LBE and ADE. The long-time behavior is shown in Fig.~\ref{fig:run2_log}, in which $1-C/C_0$ is plotted.

\clearpage

\subsection{Column experiments with adsorbed solute}

%Tracer particles in underground water are often adsorbed by rock and gravel. 
Here we use adsorptive solute (zinc solution). The filling material is the standard sand (Tohoku silica sand No. 4, Kitanihon Sangyo). The median diameter of the sand is 750$\mu$m. As a preparation, we eliminate the organic matter that may have been contained in the sand by soaking it in ${\rm HNO}_3$ solution. The zinc solution is 2 ppm. We set a filter on the top of the sand bed made with glass wool. We also put the same filter at the bottom of the column.

A short column of length $12.0\,{\rm cm}$ with internal diameter $3.1\,{\rm cm}$ was used. 
%The outlet at the column end has a funnel shape in order to prevent stagnation of effluent. 
We perform a blank test beforehand to make sure that the zinc is not absorbed on the surface of the column wall. The concentration is measured with the atomic absorption photometer (Z-2300, Hitachi High-Technologies), the compressor (SC820, Koki Holdings), and the neo cool circulator (CF700, Yamato).

\subsubsection{Run 3}
\label{exp_run3}

The measured porosity is $0.295$ and the flow rate is $9.34\,{\rm cm}^3/{\rm h}$. The bed height is $10.7\,{\rm cm}$ and the section area is $7.54\,{\rm cm}^2$.

\begin{figure}
\centering
\includegraphics[clip,width=0.8\textwidth]{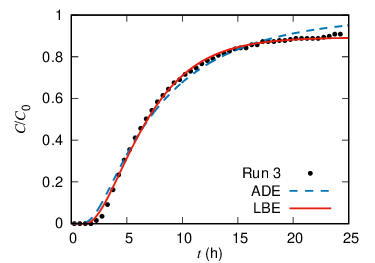}
\caption{
The relative concentration $C/C_0$ for Run 3 (black dots) is plotted with calculated values by LBE (red solid line) and by ADE (blue dashed line).
}
\label{fig:run3}
\end{figure}

\begin{figure}
\centering
\includegraphics[clip,width=0.8\textwidth]{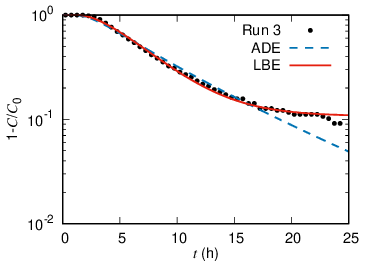}
\caption{
Same as Fig.~\ref{fig:run3} but $1-C/C_0$ is plotted in the semi-log graph.
}
\label{fig:run3_log}
\end{figure}

Figure \ref{fig:run3} shows the relative concentration $C/C_0$ of the zinc breakthrough curve for Run 3 (black dots) with calculated values from the linear Boltzmann equation (LBE) (red solid line) and the Ogata-Banks ADE model (blue dashed line). By nonlinear least-squares fitting, we obtain $\sigma_a=1.0\times 10^{-7}\,{\rm h}^{-1}$, $\sigma_s=2.8134\,{\rm h}^{-1}$, $v_0=5.0663\,{\rm cm}/{\rm h}$, $u=1.9876\,{\rm cm}/{\rm h}$, $\beta=0.1739$ for LBE, and $u=1.1406\,{\rm cm}/{\rm h}$, $D=4.3864\,{\rm cm}^2/{\rm h}$. The dashed curve for ADE partially deviates from the measured values. The semi-log plot in Fig.~\ref{fig:run3_log} shows a clear discrepancy between the experimental result and ADE.

\subsubsection{Run 4}
\label{exp_run4}

\begin{figure}
\centering
\includegraphics[clip,width=0.8\textwidth]{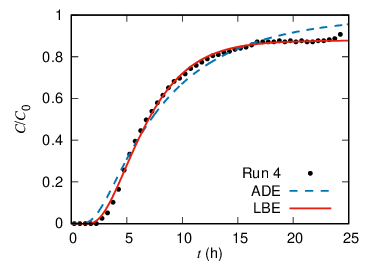}
\caption{
The relative concentration $C/C_0$ for Run 4 (black dots) is plotted with calculated values by LBE (red solid line) and by ADE (blue dashed line).
}
\label{fig:run4}
\end{figure}

\begin{figure}
\centering
\includegraphics[clip,width=0.8\textwidth]{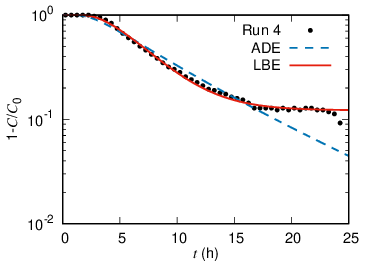}
\caption{
Same as Fig.~\ref{fig:run4} but $1-C/C_0$ is plotted in the semi-log graph.
}
\label{fig:run4_log}
\end{figure}

Figure \ref{fig:run4} for Run 4 is the same as Fig.~\ref{fig:run3} but we repeat another run for the standard sand and adsorbed solute. The measured porosity is $0.289$ and the flow rate is $11.25\,{\rm cm}^3/{\rm h}$. The bed height is $10.7\,{\rm cm}$ and the section area is $7.54\,{\rm cm}^2$. By nonlinear least-squares fitting, we obtain $\sigma_a=1.0\times 10^{-7}\,{\rm h}^{-1}$, $\sigma_s=2.6773\,{\rm h}^{-1}$, $v_0=4.1166\,{\rm cm}/{\rm h}$, $u=1.7999\,{\rm cm}/{\rm h}$, $\beta=0.2091$ for LBE, and $u=1.1384\,{\rm cm}/{\rm h}$, $D=3.9814\,{\rm cm}^2/{\rm h}$ for ADE. The discrepancy of ADE is even more apparent. The semi-log plot in Fig.~\ref{fig:run4_log} shows a discrepancy between the experimental result and ADE.

\clearpage
\section{Conclusions}
\label{sec:concl}

Not only for field observation reported in studies such as Adams and Gelhar (\citealt{Adams_1992}) and Benson and his collaborators (\citealt{Benson_2001,Benson_2000}), anomalous diffusion also appears in lab-scale experiments (\citealt{Cotris_2004}). In Sec.~\ref{sec:column}, ADE does not precisely reproduce the breakthrough curves even for beads, which form a medium material with non-adsorbed solution. Williams proposed to use the linear Boltzmann equation for transport in porous media (\citealt{Williams_1992a,Williams_1992b,Williams_1993a,Williams_1993b}). It was shown that the proposed model of the linear Boltzmann equation reproduces the whole breakthrough curves of experimental results including the non-monotonic decay of $1-C/C_0$ in the tail.

The diffusion approximation holds when the propagation distance of tracer particles is sufficiently larger than the transport mean free path $\ell^*$, which is given in (\ref{LBE_ADE_9}). In the case of light transport, the photon diffusion equation starts to work when the propagation distance becomes $10$ times larger than the transport mean free path (\citealt{Yoo_1990}). Using the relation (\ref{LBE_ADE_8}), we can calculate the diffusion coefficient, which is denoted by $D'$, from parameters in the linear Boltzmann equation. We let $D$ denote the fitted diffusion coefficient in the advection-diffusion equation. Let us compare Run 1 and Run 2. For Run 1 in Sec.~\ref{exp_run1}, $\ell^*=1.03$ and $D'=1.818$, whereas $D=1.8379$. The relative difference is $|D-D'|/D=0.0108$. For Run 2 in Sec.~\ref{exp_run2}, $\ell^*=1.75$ and $D'=4.762$, whereas $D=3.8519$. The relative difference is $|D-D'|/D=0.236$. In either case, the propagation distance of $18\,{\rm cm}$ or $18.5\,{\rm cm}$ is much larger than $\ell^*$ and the transport is expected to be in the diffusion regime. Indeed, Fig.~\ref{fig:run1} and Fig.~\ref{fig:run2} show that both LBE and ADE reproduce the experimentally obtained breakthrough curves. Next, let us also look at Run 3 and Run 4. For Run 3 in Sec.~\ref{exp_run3}, $\ell^*=1.80$ and $D'=3.041$, whereas $D=4.3864$. The relative difference is $|D-D'|/D=0.307$. For Run 4 in Sec.~\ref{exp_run4}, $\ell^*=1.54$ and $D'=2.110$, whereas $D=3.9814$. The relative difference is $|D-D'|/D=0.470$. The propagation distance of $10.7\,{\rm cm}$ is not as long as the distance in Run 1 and Run 2 in terms of the transport mean free path. The discrepancy of the curve for ADE in Fig.~\ref{fig:run3} and Fig.~\ref{fig:run4} might also attribute to anisotropic scattering for the sand. The transport mean free path becomes larger when the scattering in LBE becomes anisotropic. Moreover the discrepancy may come from the fact that the surface of the sand is rough and the particle speed cannot be modeled by a constant $v_0$.
 
Although we assumed isotropic scattering in (\ref{proposed_model}), it is possible to take anisotropic scattering into account and write the integral term in a more general form as $\sigma_s\int_{-1}^1p(\mu,\mu')\psi(x,\mu',t)\,\mathrm{d}\mu'$, where $p(\mu,\mu')$ is the probability that tracer particles moving in the direction $\mu'$ change their direction to the direction $\mu$ by scattering. We can similarly develop the analytical discrete ordinates method for anisotropic scattering by writing $p(\mu,\mu')=\frac{1}{2}\sum_l\beta_lP_l(\mu)P_l(\mu')$ with some constant coefficients $\beta_l$ and Legendre polynomials $P_l$ ($l=0,1,\dots$) (\citealt{Barichello_2011,Siewert_2000}).

In this paper we have shown that the mass transport in column experiments, which show anomalous diffusion, obeys the linear Boltzmann equation. Furthermore the linear Boltzmann equation in the transport regime appears at the mesoscopic scale and the advection-diffusion equation in the diffusion regime is derived from the linear Boltzmann equation in the asymptotic limit when the propagation distance is sufficiently larger than $\ell^*$ at the macroscopic scale. The physical origin of anomalous diffusion in a heterogeneous random medium with more complex structure of fissures is still an open problem.

\begin{acknowledgements}
The seed of this research was planted on the occasion of the Study Group Workshop (Department of Mathematical Sciences, The University of Tokyo, December 2010), which is greatly appreciated. The research was restarted with the support of the Focusing Collaborative Research by the Interdisciplinary Project on Environmental Transfer of Radionuclides (University of Tsukuba and Hirosaki University). The research was partially supported by the JSPS A3 foresight program: Modeling and Computation of Applied Inverse Problems. MM acknowledges support from Grant-in-Aid for Scientific Research (17H02081,17K05572) of the Japan Society for the Promotion of Science. YH acknowledges support from Grant-in-Aid for Scientific Research (15H05740,17H01478) of the Japan Society for the Promotion of Science. 
\end{acknowledgements}

\appendix

\section{Eigenvalues}
\label{eigen}

Recall $\mu_t(p)=(\sigma_a+\sigma_s+p)/v_0$, $\mu_s=\sigma_s/v_0$ with positive constants $\sigma_a,\sigma_s,v_0$ ($p$ is complex). Let us write the homogeneous equation as
\begin{equation}
\begin{aligned}
\left(\mu_t(p)I-\frac{1}{\nu}\Xi_+\right)\Phi_+
&=
\frac{\mu_s}{2}W\left(\Phi_++\Phi_-\right),
\\
\left(\mu_t(p)I-\frac{1}{\nu}\Xi_-\right)\Phi_-
&=
\frac{\mu_s}{2}W\left(\Phi_++\Phi_-\right),
\end{aligned}
\end{equation}
where the $N\times N$ matrix $I$ is the identity, $\Xi_{\pm}=\mathop{\mathrm{diag}}(\pm\mu_1+\eta,\dots,\pm\mu_N+\eta)$. We defined $W$ and $\Phi_{\pm}$ as
\begin{equation}
\Phi_{\pm}=D\begin{pmatrix}
\phi(\nu,\pm\mu_1) \\ \phi(\nu,\pm\mu_2) \\ \vdots \\ \phi(\nu,\pm\mu_N)
\end{pmatrix},
\quad
\left\{W\right\}_{ij}=\sqrt{w_iw_j},
\end{equation}
where $D=\mathop{\mathrm{diag}}(\sqrt{w_1},\dots,\sqrt{w_N})$. We arrive at
\begin{equation}
SM(p)\varphi=\frac{1}{\nu}\varphi.
\label{eigenprob}
\end{equation}
Here, the diagonal matrix $S$ and symmetric matrix $M(p)$ are given by
\begin{equation}
S=\begin{pmatrix}\Xi_+^{-1} & \\ & \Xi_-^{-1}\end{pmatrix},
\quad
M(p)=\mu_t(p)I_{2N}-\frac{\mu_s}{2}\begin{pmatrix}W & W \\ W & W\end{pmatrix},
\quad
\varphi=\begin{pmatrix}\Phi_+ \\ \Phi_-\end{pmatrix},
\quad
I_{2N}=\begin{pmatrix}I & \\ & I \end{pmatrix}.
\end{equation}

We can write
\begin{equation}
\frac{1}{2}\begin{pmatrix}W & W \\ W & W\end{pmatrix}=\xi\xi^T,
\quad
\xi=\frac{1}{\sqrt{2}}\left(\sqrt{w_1}\cdots\sqrt{w_N}\;\sqrt{w_1}\cdots\sqrt{w_N}\right)^T,
\end{equation}
where $T$ means transpose. We note $|\xi|=1$. Similarly, we can write $W=vv^T$ ($v_i=\sqrt{w_i}$).

For any $N$-dimensional complex vector $x$, $Wx=vv^Tx=(v\cdot x)v$. That is, $Wx$ is in the diretion of $v$. Hence $W$ is rank $1$ ($\mathop{\mathrm{rank}}(W)=\mathop{\mathrm{dim}}(\mathop{\mathrm{Im}}(W))=1$) and $v$ is an eigenvector ($Wv=|v|^2v$). For any vector $v_{\perp}$ ($v\cdot v_{\perp}=0$), $Wv_{\perp}=v(v\cdot v_{\perp})=0$. Hence, there is one eigenvalue $|v|^2=\sum_{i=1}^Nw_i=1$ and $N-1$ zero eigenvalues. Noting that $\begin{pmatrix}1&1\\1&1\end{pmatrix}$ has eigenvalues $2,0$, we see that the eigenvalues of $\begin{pmatrix}1&1\\1&1\end{pmatrix}\otimes W$ are one $2$ and $2N-1$ zeros (the eigenvalues of the Kronecker product of two matrices are the pairwise products of the eigenvalues of the two matrices). Hence, in the eigenvalues of $M(p)$, $\mu_t(p)-\mu_s$ appears once and $\mu_t(p)$ appears $2N-1$ times. The real parts of both eigenvalues are positive when $\Re p>-\sigma_a$. If $p$ is real with $p>-\sigma_a$, then $M$ is positive definite because all eigenvalues of $M$ are positive. 

Let us suppose $p$ is real and $p>-\sigma_a$. We write $S=\Sigma A$, where $\Sigma=\mathop{\mathrm{sign}}(S)$ (each diagonal element of $\Sigma$ is $1$ or $-1$ depending on the corresponding diagonal element of $S$) and $A$ is a positive definite diagonal matrix. Define $K=A^{1/2}MA^{1/2}$. We note that $K^{1/2}$ is symmetric, real, and positive definite ($K^{1/2}:=Q\Theta^{1/2}Q^T$, $\Theta^{1/2}=\mathop{\mathrm{diag}}(\sqrt{\theta_1},\dots,\sqrt{\theta_{2N}})$, where $Q^TQ=QQ^T=I_{2N}$, $\Theta=\mathop{\mathrm{diag}}(\theta_1,\dots,\theta_{2N})$). First, we notice that the eigenvalues of $(A^{1/2}K^{-1/2})^{-1}SM(A^{1/2}K^{-1/2})$ are the same as the eigenvalues of $SM$ because these two matrices are similar. Second, we have
\begin{equation}
K^{1/2}A^{-1/2}SMA^{1/2}K^{-1/2}
=K^{1/2}\Sigma(A^{1/2}MA^{1/2})K^{-1/2}
=K^{1/2}\Sigma K^{1/2}.
\end{equation}
Since $K^{1/2}$ is nonsingular and $\Sigma$ is symmetric, matrices $K^{1/2}\Sigma K^{1/2}$ and $\Sigma$ have the same inertia, i.e., the same number of positive, negative, and zero eigenvalues according to Sylvester's law of inertia. Since the number of $1$ in $\Sigma$ is $N_{\eta}$, we see that $SM$ has $N_{\eta}$ positive eigenvalues. Let $\nu_0$ be an eigenvalue of $SM$ in this case.

Next, we suppose that $p$ is complex with $\Re p>-\sigma_a$ and consider the eigenproblem (\ref{eigenprob}). Since $\det(SM(p)-\lambda I_{2N})=0$, the complex number $\nu^{-1}=\lambda$ varies continuously with $p$ (continuity of the roots of a polynomial). By taking the product with $\varphi^*$ (conjugate transpose), we have
\begin{equation}
\varphi^*M\varphi=\lambda\varphi^*S^{-1}\varphi.
\end{equation}
We note that $\varphi^*S^{-1}\varphi$ is real and
\begin{align}
\varphi^*M\varphi
&=
\mu_t(p)|\varphi|^2-\mu_s|\xi^T\varphi|^2
\nonumber \\
&=
\frac{|\varphi|^2}{v_0}\left(\sigma_a+\sigma_s+p-\frac{|\xi^T\varphi|^2}{|\varphi|^2}\sigma_s\right).
\end{align}
Since $|\xi^T\varphi|^2/|\varphi|^2\in[0,1]$, we have
\begin{equation}
\left(\Re\lambda\right)\varphi^*S^{-1}\varphi\ge
\frac{|\varphi|^2}{v_0}\left(\sigma_a+\Re p\right)>0.
\end{equation}
This implies $\Re\lambda\neq0$. Put $\nu=\lambda^{-1}$ (delete this). Hence, if $\nu_0>0$ (or $<0$), then $\Re\nu>0$ (or $<0$) when $p$ varies continuously. The number $N_{\eta}$ of eigenvalues with $\nu_0>0$ is the number of eigenvalues with $\Re\nu>0$. In particular, this relation holds for the Bromwich contour $\Re p=\gamma$.

\section{Advection-diffusion equation with absorption}
\label{wabsorption}

Let us consider the following advection-diffusion equation with the absorption term.
\begin{align}
\left(\frac{\partial}{\partial t}-D\frac{\partial^2}{\partial x^2}+u\frac{\partial}{\partial x}+\sigma_a
\right)C(x,t)=0,
\\
C(x,0)=0,
\\
C(0,t)=C_0.
\end{align}
Let us introduce $\Gamma(x,t)$ as
\begin{equation}
C(x,t)=\Gamma(x,t)\exp\left(\frac{ux}{2D}-\frac{u^2t}{4D}-\sigma_at\right).
\end{equation}
We have
\begin{align}
\frac{\partial}{\partial t}\Gamma(x,t)-D\frac{\partial^2}{\partial x^2}\Gamma(x,t)=0,
\\
\Gamma(x,0)=0,
\\
\Gamma(0,t)=C_0\exp\left(\frac{u^2t}{4D}+\sigma_at\right).
\end{align}
Let us consider the Laplace transform:
\begin{equation}
\hat{\Gamma}(x,p)=\int_0^{\infty}\mathrm{e}^{-pt}\Gamma(x,t)\,\mathrm{d}t,
\quad\Re{p}>\frac{u^2}{4D}+\sigma_a.
\end{equation}
Then we obtain
\begin{equation}
\frac{\partial^2}{\partial x^2}\hat{\Gamma}(x,p)=\frac{p}{D}\hat{\Gamma}(x,p),\quad
\hat{\Gamma}(0,p)=\frac{C_0}{p-\frac{u^2}{4D}-\sigma_a}.
\end{equation}
We obtain
\begin{equation}
\hat{\Gamma}(x,p)=\frac{C_0}{p-\frac{u^2}{4D}-\sigma_a}
\mathrm{e}^{-\sqrt{\frac{p}{D}}x}.
\end{equation}
Thus,
\begin{equation}
\Gamma(x,t)=C_0\int_0^tf(t-\tau)g(\tau)\,d\tau,
\end{equation}
where
\begin{equation}
\hat{f}(p)=\frac{1}{p-\frac{u^2}{4D}-\sigma_a},\quad
\hat{g}(p)=\mathrm{e}^{-\sqrt{\frac{p}{D}}x}.
\end{equation}
Noting that
\begin{equation}
\int_0^{\infty}\mathrm{e}^{-pt}e^{at}\,\mathrm{d}t=
\frac{1}{p-a}\;(\Re{p}>a),\quad
\int_0^{\infty}\mathrm{e}^{-pt}\frac{b}{2\sqrt{\pi}t^{3/2}}
e^{-\frac{b^2}{4t}}\,\mathrm{d}t=
\mathrm{e}^{-b\sqrt{p}}\;(\Re{p}>0),
\end{equation}
we obtain
\begin{align}
\Gamma(x,t)
&=
C_0\int_0^t\exp\left(\left(\frac{u^2}{4D}+\sigma_a\right)(t-\tau)\right)
\frac{x/\sqrt{D}}{2\sqrt{\pi}\tau^{3/2}}\mathrm{e}^{-\frac{x^2}{4D\tau}}
\,\mathrm{d}\tau
\nonumber \\
&=
\frac{x/\sqrt{D}}{2\sqrt{\pi}}\exp\left(\left(\frac{u^2}{4D}+\sigma_a\right)t\right)
\exp\left(-\sqrt{\frac{u^2}{4D}+\sigma_a}\frac{x}{\sqrt{D}}\right)
\nonumber \\
&\times
\int_0^t\frac{1}{\tau^{3/2}}\exp\left(-\left(\frac{x/\sqrt{D}}{2\sqrt{\tau}}-
\sqrt{\left(\frac{u^2}{4D}+\sigma_a\right)\tau}\right)^2\right)
\,\mathrm{d}\tau
\nonumber \\
&=
\frac{C_0}{2}\exp\left(\left(\frac{u^2}{4D}+\sigma_a\right)t\right)
\exp\left(-\sqrt{\frac{u^2}{4D}+\sigma_a}\frac{x}{\sqrt{D}}\right)
\nonumber \\
&\times
\Biggl[
\mathop{\mathrm{erfc}}\left(\frac{x}{\sqrt{4Dt}}-\sqrt{\left(\frac{u^2}{4D}+\sigma_a\right)t}\right)
\nonumber \\
&+
\exp\left(\sqrt{\frac{u^2}{4D}+\sigma_a}\frac{2x}{\sqrt{D}}\right)
\mathop{\mathrm{erfc}}\left(\frac{x}{\sqrt{4Dt}}+\sqrt{\left(\frac{u^2}{4D}+\sigma_a\right)t}\right)
\Biggr],
\end{align}
where the complementary error function is given by $\mathop{\mathrm{erfc}}(x)=(2/\sqrt{\pi})\int_x^{\infty}\mathrm{e}^{-t^2}\,\mathrm{d}t$. Finally,
\begin{equation}
\begin{aligned}
C(x,t)
&=
\frac{C_0}{2}\mathrm{e}^{\frac{ux}{2D}}
\exp\left(-\sqrt{\frac{u^2}{4D}+\sigma_a}\frac{x}{\sqrt{D}}\right)
\\
&\times
\Biggl[
\mathop{\mathrm{erfc}}\left(\frac{x}{\sqrt{4Dt}}-\sqrt{\left(\frac{u^2}{4D}+\sigma_a\right)t}\right)
\\
&+
\exp\left(\sqrt{\frac{u^2}{4D}+\sigma_a}\frac{2x}{\sqrt{D}}\right)
\mathop{\mathrm{erfc}}\left(\frac{x}{\sqrt{4Dt}}+\sqrt{\left(\frac{u^2}{4D}+\sigma_a\right)t}\right)
\Biggr],
\end{aligned}
\end{equation}
The above solution reduces to the Ogata-Banks solution (\citealt{Ogata_1961}) when $\sigma_a=0$. In particular, we have $C(x,t)\to C_0$ as $t\to\infty$ if $\sigma_a=0$. When $\sigma_a=0$, we have
\begin{equation}
\label{appendix:ogata-banks}
C(x,t)=
\frac{C_0}{2}\left[
\mathop{\mathrm{erfc}}\left(\frac{x-ut}{\sqrt{4Dt}}\right)+
\mathrm{e}^{ux/D}
\mathop{\mathrm{erfc}}\left(\frac{x+ut}{\sqrt{4Dt}}\right)
\right].
\end{equation}



\end{document}